\documentclass{iopart}
\usepackage{iopams}
\usepackage{setstack}

\usepackage{graphicx}
\usepackage{color}

\begin{document}

\topical[The role of phonons for exciton and biexciton generation] {The role
of phonons for exciton and biexciton generation in an optically driven
quantum dot}
\author{D~E~Reiter$^1$, T~Kuhn$^1$, M~Gl\"assl$^2$, V~M~Axt$^2$}
\address{$^1$ Institut f\"ur Festk\"orpertheorie, Universit\"at M\"unster, Wilhelm-Klemm-Stra\ss{}e 10, 48149 M\"unster, Germany}
\address{$^2$ Theoretische Physik III, Universit\"at Bayreuth, 95440 Bayreuth, Germany}

\begin{abstract}
For many applications of semiconductor quantum dots in quantum technology a
well controlled state preparation of the quantum dot states is mandatory.
Since quantum dots are embedded in the semiconductor matrix, the interaction
with phonons plays often a major role in the preparation process. In this
review, we discuss the influence of phonons on three basically different
optical excitation schemes which can be used for the preparation of exciton,
biexciton, and superposition states: a resonant excitation leading to Rabi
rotations in the excitonic system, an excitation with chirped pulses
exploiting the effect of adiabatic rapid passage, and an off-resonant
excitation giving rise to a phonon-assisted state preparation. We give an
overview over experimental and theoretical results showing the role of the
phonons and compare the performance of the schemes for state preparation.
\end{abstract}



\section{Introduction}
\label{sec:introduction}

Quantum dots (QDs) are semiconductor nanostructures that combine aspects
typical for atomic systems, in particular a discrete energy spectrum, with
the technologically well developed and well controllable semiconductor
material system. This compatibility with existing technologies makes QDs
attractive candidates for a wide range of applications ranging from
optoelectronic devices like QD lasers
\cite{bimberg1999qua,schweizer2003qua,ledentsov2011qua,chow2013phy}, where
the target is an improved performance compared to other laser structures, up
to new applications in the fields of quantum cryptography or quantum
information processing, where the presence of discrete energy levels is
mandatory. Examples of such quantum applications are single-photon sources
\cite{michler2000qua,santori2001tri,yuan2002ele,press2007pho,yilmaz2010qua,eisaman2011inv},
sources of entangled photon pairs
\cite{benson2000reg,moreau2001qua,santori2002pol,akopian2006ent,stevenson2006ase,hafenbrak2007tri,dousse2010ult,muller2014ond},
and qubit devices or quantum gates
\cite{bonadeo1998coh,biolatti2000qua,troiani2000exp,damico2002all,boyle2008two,michaelis2010coh}.
The functionality of all these latter applications relies on the preparation
of a well-defined quantum state. The generation of single photons or
entangled photon pairs is based on the radiative recombination starting from
the exciton or the biexciton state. Therefore, for a highly efficient photon
generation a high-fidelity preparation of the QD in the exciton or biexciton
state is required. For the realization of a qubit or a quantum gate in
addition arbitrary superposition states have to be prepared with high
fidelity.

In an ideal optically driven two- or few-level system a preparation of
arbitrary quantum states provides no conceptual difficulty and can be
achieved, e.g., by using various coherent control techniques
\cite{ramsay2010are}. A QD, however, due to its embedding in a semiconductor
matrix and the resulting non-negligible coupling to the phonons of the
crystal lattice, is not an ideal few-level system. Phonons lead to a
dephasing of coherences and to relaxation processes between different
electronic states. This may prohibit some of the preparation schemes or at
least strongly limit the range of excitation parameters where a high-fidelity
state preparation is possible. Therefore, a detailed knowledge of the role of
phonons in the various excitation schemes is necessary to select the best
scheme and to optimize its performance.

In this review we will give an overview of the role of phonons in different
state preparation schemes using optical excitation of a single QD. In
particular, we will discuss the phonon influence on coherent excitation
schemes based on resonant excitation or excitation by chirped laser pulses,
where phonons typically give rise to unwanted dephasing or relaxation effects
and therefore introduce limitations to the otherwise ideal schemes. In
contrast to these coherent schemes, phonons may also be actively used in
state preparation by employing an excitation with detuned pulses giving rise
to a phonon-assisted exciton or biexciton generation.

The review is organized as follows. In Sec.~\ref{sec:model} the model for the
description of the electronic structure of the QD as well as its coupling to
light and phonons is introduced. The relevant physics and the different
theoretical approaches which have been used to model the dynamics of the
optically driven QD are briefly explained. In Sec.~\ref{sec:schemes} the
three basically different state preparation schemes are described.
Experimental and theoretical results that have been obtained with these
schemes are discussed in particular in view of the role of the phonons for
the state preparation. In Sec.~\ref{sec:conclusions} we compare the three
schemes and discuss their respective advantages, drawbacks, and limitations.
The review then finishes with some concluding remarks.

\section{Theoretical background}
\label{sec:model}

\subsection{Quantum dot model}
\label{sec:qdmodel}

In QDs the electronic motion is confined in all three directions on a
nanometer scale \cite{bimberg1999qua,rossi2005sem,jacak1998qua}. Therefore,
the corresponding electronic spectra possess a discrete part similar to what
is found for atoms. The target states of the preparation schemes to be
discussed in this review are typically well separated in energy from all
other electronic states, which allows for a selective excitation of these
states. This condition is best fulfilled for strongly confined dots. For the
theory, this implies that one can concentrate on a rather limited electronic
subspace spanned by states $|\nu\rangle$ with discrete energies
$\hbar\omega_{\nu}$ such that the electronic Hamiltonian reads:
\begin{eqnarray}
 \label{eq:Hel}
 H_{\mathrm{dot}}
=\sum_{\nu} \hbar\omega_{\nu}|\nu\rangle\langle\nu|.
\end{eqnarray}
Most often discussed is the case, where the valence band states are formed
from heavy-hole states with angular momentum projections $J_z=\pm 3/2$ while
the conduction band states have $J_z=\pm 1/2$. For strongly confined dots,
one can concentrate on the lowest lying electron and hole states. In this
review we will restrict ourselves to the case of charge-neutral QDs, which
implies that one has to consider a basis of four electronic states, namely
the ground state $|G\rangle$ (i.e., the state without electron-hole pairs),
two bright single exciton states $|X_{\pm}\rangle$ corresponding to single
electron-hole pairs with total angular momenta $\pm 1$ and the biexciton
state $|B\rangle$, where two electron-hole pairs are confined in the dot. Two
further exciton states with angular momenta $\pm 2$ can be formed from the
lowest single particle states, which are usually referred to as dark excitons
as they cannot be excited directly by the laser field. A relaxation into
these dark states requires spin flip processes that take place on a time
scale of typically longer than a nanosecond
\cite{paillard2001spi,tsitsishvili2005exc,roszak2007exc} or relaxation from
energetically higher excited states \cite{poem2010acc}. Therefore, they are
not relevant for state preparation schemes occurring on a time scale of at
most a few tens of picoseconds and will not be considered in this review.

In most real QDs the two bright exciton states exhibit an additional coupling
due to the long-range exchange interaction:
\begin{eqnarray}
 H_{\mathrm{exchange}}
= V_{\mathrm{ex}} \big(
|X_{+}\rangle\langle X_{-}| +
|X_{-}\rangle\langle X_{+}|
\big).
\end{eqnarray}
A finite $V_{\mathrm{ex}}$ entails a splitting of the exciton energies by
$2|V_{\mathrm{ex}}|$ and the exciton eigenstates are then given by the
linearly polarized single exciton states:
\begin{eqnarray}
|X_x\rangle = \frac{1}{\sqrt{2}}\big(|X_{+}\rangle+|X_{-}\rangle\big),
\qquad
|X_y\rangle = \frac{i}{\sqrt{2}}\big(|X_{+}\rangle-|X_{-}\rangle\big).
\end{eqnarray}
The size of the exchange splitting strongly depends on the dot geometry.
Typical values in InGaAs QDs range from almost zero to a few $100$~$\mu$eV
\cite{langbein2004con,seguin2005siz}. For schemes that are targeted at the
preparation of the biexciton state, in order to subsequently initiate a decay
cascade that creates entangled photon pairs
\cite{moreau2001qua,stevenson2006ase,dousse2010ult} it is a necessary
precondition to ideally have a vanishing exchange splitting
\cite{stevenson2006ase,hafenbrak2007tri,ghali2012gen,kuroda2013sym}.
Otherwise, a kind of which-path information would be introduced in the decay
that prevents a high degree of entanglement
\cite{akopian2006ent,coish2009ent}. Apart from selecting QDs that happen to
exhibit almost zero splitting \cite{hafenbrak2007tri}, also specially
designed growth \cite{nicoll2009mbe} or annealing \cite{tartakovskii2004eff}
techniques have been developed  for that purpose. Alternatively, one can use
dots with high symmetry such as, e.g., self-organized In(Ga)As/GaAs QDs grown
on (111) substrate which should ideally have no exchange splitting
\cite{schliwa2009ing}, or apply strategies to actively suppress the
splitting, such as, e.g., the ac-Stark tuning \cite{muller2009cre} or
application of strain or external electric and/or magnetic fields
\cite{bayer2002fin,kowalik2005inf,seidl2006eff,stevenson2006mag,trotta2012uni,welander2012ele}.
Furthermore, all preparation schemes aim at being as fast as possible without
degrading other properties such as the robustness or the spectral selectivity
of the scheme. With the schemes discussed in this review the preparation is
typically completed after about $4-20$~ps. Together with the fact that
preparation schemes are preferably executed with dots having a rather small
exchange splitting, it is justified to ignore the impact of the exchange
interaction on the time scale required for the preparation. Therefore, in the
following we shall identify  the relevant dot eigenstates with the states:
$|G\rangle$, $|X_{\pm}\rangle$ and $|B\rangle$, i.e., we have two
energetically degenerate single exciton states with energy $\hbar
\omega_{X}$. When the energy of the ground state is set to zero, the energy
of the biexciton state $\hbar \omega_{B}$ is given by $\hbar \omega_{B}= 2
\hbar \omega_{X}-\Delta_{B}$, where $\Delta_{B}$ denotes the biexciton
binding energy.

\begin{figure}[ht!]
\centering
\includegraphics[width=12cm]{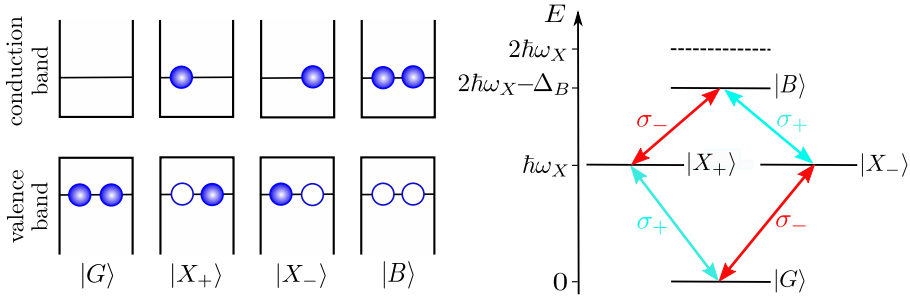}
\caption{\label{fig:level}
Electronic levels and excitation-level diagram for an exciton-biexciton QD
system in the case of a vanishing exchange interaction. $|G\rangle$ denotes the
ground state without electron-hole pairs, $|X_{\pm}\rangle$ are two single
exciton states of different angular momentum, which can be excited separately by
$\sigma^{\pm}$ circularly polarized light, and $|B\rangle$ is the biexciton
state. In the left figure, filled circles represent electrons, whereas open
circles represent holes. In the right figure, $\hbar \omega_{X}$ denotes the
ground state to exciton transition energy and $\Delta_B$ is the biexciton
binding energy. }
\end{figure}

The external driving of the system by a coherent laser field is commonly
modeled by the dipole coupling to a classical light field in the rotating
wave approximation corresponding to the Hamiltonian:
\begin{eqnarray}
 \label{eq:Hdip}
 H_{\mathrm{dot-laser}}
= -\sum_{\nu\nu'} \hbar M_{\nu\nu'}|\nu\rangle\langle\nu'|,
\end{eqnarray}
where $M_{\nu\nu^{\prime}}$ is the matrix of dipole interactions. Accounting
for the usual dipole selection rules between the dot basis states, the dipole
interaction matrix $M$ reads:
\begin{eqnarray}
M =\frac{1}{2} \left(
\begin{array}{cccc}
0      & \Omega^*_{\sigma^+}(t) & \Omega^*_{\sigma^-}(t) & 0\\
\Omega_{\sigma^+}(t)  & 0                   & 0                   & \Omega^*_{\sigma^-}(t)\\
\Omega_{\sigma^-} (t) & 0                   & 0                   & \Omega^*_{\sigma^+}(t)\\
0                  & \Omega_{\sigma^-}(t)   & \Omega_{\sigma^+}(t)   & 0
\end{array}
\right),
\label{eq:Mmat}
\end{eqnarray}
where  $\Omega_{\sigma^\pm}(t)\equiv f_{\sigma^{\pm}}(t) e^{-i\omega_{L}t}$
with $f_{\sigma^{\pm}}(t)=2M_0 E_{\sigma^\pm}(t)\,/\hbar$. Here,
$E_{\sigma^\pm}(t)$ is the circularly $\sigma^\pm$ polarized component of the
laser field which drives the dipole transitions between the ground state
$|G\rangle$ and the single exciton states $|X_{\pm}\rangle$  as well as from
the single exciton states $|X_{\mp}\rangle$ to the biexciton state
$|B\rangle$, $M_{0}$ denotes the corresponding dipole matrix element,
$\omega_{L}$ is the central frequency of the laser, and $f_{\sigma^{\pm}}(t)$
is the driving amplitude of the corresponding transition, which is defined in
such a way that for a resonant cw excitation its modulus corresponds to the
Rabi frequency. This coupling scheme is illustrated in Fig.~\ref{fig:level}.

In many studies, this model is further reduced to an electronic two-level
system (TLS) by considering an excitation by light with a single circular
polarization. For this choice of excitation conditions, two of the four
electronic states introduced above are decoupled from the dynamics and the
electronic system is reduced to the TLS formed by the ground state
$|G\rangle$ and the single exciton state that we shall denote by $|X\rangle$
driven by a light field with the amplitude $f$.

For laser pulse envelopes that are varying sufficiently slowly in time, the
dressed state basis for the electronic state space is a natural choice, which
also helps to highlight the physics of some of the preparation schemes
discussed in Sec.~\ref{sec:schemes}. These dressed states are defined as the
instantaneous eigenstates of the light-matter Hamiltonian
$H_{\mathrm{dot}}+H_{\mathrm{dot-laser}}$ in the frame rotating with the
laser frequency. The simplest case is constituted by the TLS, where the
dressed states $|D_{\pm}\rangle$ are given by:
\begin{equation}
|D_{\pm}\rangle = N_{\pm} \left[ f^* |G\rangle + \left( \Delta
\mp \sqrt{\Delta^{2}+|f|^2} \right) |X\rangle \right],
\label{eq:dressed-states}
\end{equation}
where $\Delta$ is the detuning between the laser frequency and the transition
frequency of the dot, and $N_{\pm}$ are normalization factors. The energies
of the dressed states are
\begin{equation}
\hbar \Omega_{\pm} =\frac{\hbar}{2}  \left( -\Delta \pm \sqrt{\Delta^{2}+|f|^2} \right).
\end{equation}
As can be seen from these formulas, the dressed states depend on the detuning
$\Delta$ and the strength $f$ of the optical driving. For a four-level
exciton-biexciton system, simple formulas for the dressed states arise in the
special case of a vanishing biexciton binding energy $\Delta_{B}$ and
resonant driving \cite{glassl2012pol}. For finite $\Delta_{B}$ and/or
off-resonant driving analytical expressions can still be derived, but they
are quite involved and not very instructive.

Due to their embedding in a solid-state environment, QDs cannot be treated as
isolated systems. Here, most important is the coupling to phonons. For
strongly confined dots, on short time scales, the coupling of excitons to
acoustic phonons via the pure dephasing mechanism
\cite{mahan2000man,takagahara1999the,besombes2001aco,krummheuer2002the} has
been identified as typically being the dominant source of decoherence
\cite{besombes2001aco,vagov2004non,ramsay2010dam,ramsay2010pho}. The
corresponding Hamiltonian reads:
   \begin{equation}
\label{eq:Hph}
H_{\mathrm{dot-phonon}} \!=\! \sum_{\bf q} \hbar\omega_{\bf q}\,b^\dag_{\bf q} b_{\bf q}
\!+\! \sum_{{\bf q}, \nu} \hbar  \big( g_{\bf q}^{\nu} b_{\bf q} \!+\! g^{\nu\ast}_{\bf q} b^\dag_{\bf q}
                          \big) |\nu \rangle\langle \nu|.
\end{equation}
The operator $b^\dag_{\bf q}$ ($b_{\bf q}$) creates (annihilates) an acoustic
phonon in a mode labeled by ${\bf q}$, and $\hbar \omega_{\bf q}$ is the
corresponding phonon energy. In this review we will mainly concentrate on
strongly confined self-assembled InGaAs or GaAs QDs, where the lattice
properties in- and outside the dot are similar. Therefore, the phonon modes
can be approximated by bulk modes, where the label ${\bf q}$ denotes the
phonon wave vector. Furthermore, for strongly confined dots, the excitonic
and biexcitonic wave functions approximately factorize into products of
single particle electron and hole wave functions, which implies that the
phonon coupling constants are of the form: $g_{\bf q}^{\nu}=n_{\nu} \,
g_{{\bf q}}$, where $n_{\nu}$ represents the number of excitons present in
the state $|\nu\rangle$ and $g_{\bf q}$ is the exciton-phonon coupling
\cite{krummheuer2002the}. The explicit form of $g_{\bf q}$ depends on the
specific coupling mechanism. For typical InGaAs or GaAs dots, as discussed
mostly in this review, the deformation potential coupling to longitudinal
acoustic (LA) phonons is by far the strongest, while for strongly polar
materials like GaN or QDs with spatially strongly separated electron and hole
wave functions, the piezoelectric coupling to both longitudinal and
transverse phonons may become dominant \cite{krummheuer2002the,
krummheuer2005pur, hodgson2008dec}.

The carriers in the QD also interact with optical phonons, e.g., via the
Fr\"ohlich interaction
\cite{hameau1999str,carmele2010ant,schuh2009rab,reiter2011gen}, which under
certain conditions can also contribute to the generation of biexcitons
\cite{findeis2000pho}. In contrast to acoustic phonons, optical phonons, in
particular those which are efficiently coupled to the electrons and holes in
a QD, can typically be taken to be dispersionless. Because of their
essentially discrete spectrum, the interaction with optical phonons does not
give rise to dephasing, instead it leads to an oscillatory contribution to
the dynamics of the polarization visible as discrete sideband in the optical
spectrum \cite{krummheuer2002the,stock2011aco}. However, the interaction
strength of the energetically well separated optical phonons compared to
acoustic phonons is typically rather small and except for very short pulse
excitations they are usually out of the spectral range of excitation.
Therefore, we will focus on the acoustic phonons in the remainder of this
review. Explicit expressions for all commonly used carrier-phonon coupling
schemes for QDs can be found in Ref.~\cite{krummheuer2002the}.

\begin{figure}[ht!]
\centering
\includegraphics[width=7cm]{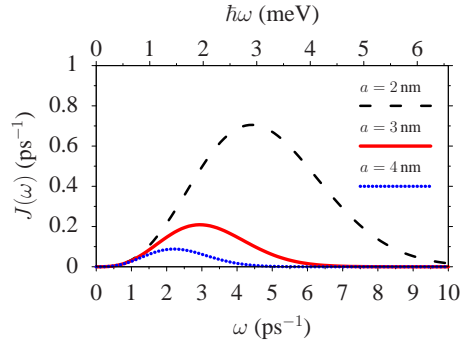}
\caption{\label{fig:spectraldensity}
Phonon spectral density for different localization lengths
of the QD wave functions.
 }
\end{figure}

Most important in the context of state preparation is the influence of the
phonon degrees of freedom on the electronic dynamics. The latter is
determined by the phonon spectral density $J(\omega)$, which is defined by:
\begin{equation}
 J(\omega) = \sum_{\bf q} |g_{\bf q}|^2 \delta(\omega-\omega_{\bf q}) \, .
\label{eq:phonon-spect}
\end{equation}
For QDs with a harmonic confinement potential in which the carriers are
coupled to LA phonons via the deformation potential the phonon spectral
density is well approximated by \cite{calarco2003spi,mccutcheon2010qua}:
\begin{equation}
 J(\omega) = A \omega^3 \exp(-\omega^2/\omega_c^2)
\label{eq:phonon-spect-anal}
\end{equation}
with a strength $A$ and a cut-off frequency $\omega_c$ determined by the size
of the QD. For a spherical QD with equal confinement lengths $a$ for
electrons and holes this formula is exact and the parameters are given by
\begin{equation}
A=\frac{|D_e-D_h|^2}{4\pi^2 \varrho \hbar c_s^5} \quad \mathrm{and} \quad \omega_c= \sqrt{2}\frac{c_s}{a} ,
\end{equation}
where $D_{e/h}$ are the deformation potentials of electrons and holes,
$\varrho$ is the crystal density, and $c_s$ is the longitudinal sound
velocity. Figure~\ref{fig:spectraldensity} shows $J(\omega)$ for three
different dot sizes and GaAs parameters. It is seen that $J(\omega)$ is
maximal at a finite phonon frequency, which is larger for smaller dots and
typically of the order of a few meV.  For small frequencies the spectral
density scales according to $J(\omega) \sim \omega^{3}$. Such a power-law
behavior with an exponent $n>1$ is referred to as superohmic coupling
\cite{weiss2008qua}.

The smooth phonon spectral density of Fig.~\ref{fig:spectraldensity} is
typical for the interaction of excitons or biexcitons in a single QD with
bulk acoustic phonon modes, where the size of the QD is the only
characteristic length scale. Additional structures in the spectral density
appear as soon as other characteristic length scales are present in the
system. Phonon couplings for various types of QD structures with more than
one length scale have been discussed in the literature, such as QDs with
different confinement lengths of electrons and holes \cite{nysteen2013pro} or
with different aspect ratios \cite{ostapenko2012exc}, strong variations of
the acoustic properties at the boundary of the QD \cite{grosse2007ele}, a QD
placed close to a surface \cite{krummheuer2005cou}, in free-standing slab
\cite{krummheuer2005cou,debald2002con} or in a quantum wire
\cite{lindwall2007zer}, or a pair of QDs at a given distance
\cite{rozbicki2008qua,gawarecki2010pho}.

The physics described by the model defined by Eqs.~(\ref{eq:Hel}),
(\ref{eq:Hdip}), and (\ref{eq:Hph}) is most easily understood by first
concentrating on two simple limiting cases: (i) vanishing exciton-phonon
coupling, i.e., $g_{\bf q}=0$, and (ii) no optical driving, i.e., $M=0$. In
the first case, the dynamics resulting from the remaining model reduces to
the well-known optical Bloch equations for electronic two- or four-level
systems without relaxation \cite{allen1975opt}. In general, the laser driving
leads to a coherent superposition of the discrete dot states, which typically
implies for the corresponding occupations an oscillatory behavior known as
Rabi oscillations. The frequency of these oscillations depends on the
characteristics of the pulse. For the simplest case of a TLS driven by a
monochromatic laser field with constant amplitude $f$ the Rabi frequency
$\Omega_{\mathrm{Rabi}}$ is given by \cite{allen1975opt}:
\begin{equation}
  \label{eq:rabi}
  \Omega_{\mathrm{Rabi}} = \sqrt{\Delta^{2}+ |f|^{2}}.
\end{equation}
In the dressed state picture these oscillations correspond to a coherent
superposition of the two dressed states resulting in quantum beats
oscillating with the difference frequency
$\Omega_{+}-\Omega_{-}=\Omega_{\mathrm{Rabi}}$. Most important for the
discussion of this review, exciton or biexciton states are reached by the
coherent Bloch-type dynamics as pure states only for special excitation
conditions. For example, in the case of a TLS the exciton can only be
prepared as a pure state for resonant excitation, i.e., for $\Delta=0$ and
distinguished values of the pulse area $\theta$, that equal odd multiples of
$\pi$. Here, the pulse area is given by:
\begin{equation}
\theta = \int_{-\infty}^{\infty} |f(t)|\, dt .
\end{equation}
The influence of phonons on the state preparation using resonant Rabi
oscillations is summarized in Sec.~\ref{sec:resonant}.

In the second limiting case the model reduces to the so called
\emph{independent Boson model} which is known to be analytically solvable
\cite{huang1950the,duke1965pho,schmitt1987the,mahan2000man}. As the
independent boson model represents a pure dephasing mechanism, i.e., there is
no coupling between different electronic states, the electronic occupations
stay constant in the course of time. However, the carrier-phonon coupling in
the independent boson model mixes electronic and phononic degrees of freedom
in the excited states and thus gives rise to a polaronic character to the
eigenstates of the system.

When both, the carrier-light and the carrier-phonon coupling are present, the
interplay of both interaction mechanisms leads to a variety of dynamical
features not found in the limiting cases, such as phonon-assisted optical
generation processes or phonon-induced modifications of Rabi oscillations, as
will be discussed in detail in Sec.~\ref{sec:schemes}. Other interaction
mechanisms than the coupling to acoustic phonons typically lead to relaxation
channels that, at least at sufficiently low temperatures, act on longer time
scales than those considered here for preparation purposes. As an example, we
mention the radiative decay, that takes place on a time scale of hundreds of
picoseconds up to nanoseconds
\cite{borri2001ult,bayer2002tem,langbein2004rad}.

For some applications, it is favorable to place the QD in a microcavity
\cite{chang2006eff,ates2009non,bruggemann2011las,kasprzak2013coh,jakubczyk2013lig,blattmann2014ent},
where the dot interacts with confined photon modes. Also for dots in cavities
phonons can play a decisive role, e.g., for the dephasing
\cite{kaer2010non,roy2011pho,roy2011inf,kaer2012mic,glassl2012int}, the
photon statistics
\cite{nazir2008pho,carmele2010ant,harsij2012inf,rastghalam2013int}, the
indistinguishability of photons \cite{kaer2013mic} as well as for providing a
dot-cavity coupling in the case of non-resonant QD and cavity modes where
phenomena like off-resonant cavity feeding have been found
\cite{naesby2008inf,ates2009non,hohenester2009pho,hohenester2010cav,calic2011pho,majumdar2011pho,hughes2011inf,ulrich2011dep}.
Although such systems are not at the focus of the present review, we note in
passing that in this case, additional relaxation mechanisms, in particular
cavity losses, may become of importance.

For more elaborate applications in quantum information technology the
question of upscaling arises. This means that systems consisting of more than
a single QD have to be considered. Like in the case of a QD in a microcavity,
also here new phonon-related aspects come into play. A detailed discussion of
such systems is beyond the scope of this review. However, let us briefly
discuss some of these new aspects. The next step in the upscaling is an
extension to double QD systems. If the QDs are sufficiently close to each
other a tunnel coupling between the electron and/or hole states of the
individual QDs may arise leading to the appearance of delocalized states.
Because of the similarity of this states with bonding or antibonding states
these structures are also named QD molecules. This coupling, which can be
controlled to a large degree by an external electric field, has been the
subject of a large amount of work over the past years investigating both
excitonic \cite{bayer2001cou,krenner2005dir,stinaff2006opt} and biexcitonic
transitions \cite{gywat2002bie,scheibner2007pho}. In particular close to a
resonance between spatially direct and indirect exciton states phonon-induced
transitions can be strongly enhanced
\cite{muljarov2005pho,gawarecki2010pho,gawarecki2012pho,daniels2013exc},
which can also be interpreted as phonon-assisted tunneling
\cite{lopez2005pho,muller2012ele}. The formation of a molecular polaron can
give rise to a phonon-induced optical transparency \cite{kerfoot2014opt}.
Even if the distance between the QDs is larger such that tunneling processes
are negligible, the QDs can still be coupled via the Coulomb interaction
which leads to an excitation transfer, often referred to as F{\"o}rster
coupling. Like the tunneling, this transfer is also affected by phonons
\cite{rozbicki2008qua}. Phonon wave packets emitted from one QD can modulate
the optical response of a second QD \cite{huneke2008imp}. Phonons in general
also give rise to the decay of entanglement between excitons in two QDs
\cite{roszak2006com}, which is an important aspect in quantum computation.

\subsection{Theoretical Methods}
\label{sec:theory}

Apart from its relevance for the laser-driven QD dynamics, the model defined
by Eqs.~(\ref{eq:Hel}), (\ref{eq:Hdip}), and (\ref{eq:Hph}) is a prototype of
a quantum dissipative system \cite{breuer2002the,weiss2008qua,leggett1987dyn}
and thus also serves as a test ground for different theoretical methods. It
is therefore not surprising that a wealth of theoretical approaches has been
actively used for exploring the resulting dynamics. Despite of the simplicity
of the electronic structure, the model in fact represents a genuine
many-particle system due to the coupling to a continuum of acoustic phonon
modes and an unlimited occupation number of each mode. For an arbitrary laser
driving no analytical solution of the model is known.

Only in the limiting case of ultrashort excitations exact analytical
solutions can be obtained, because in this limit the influence of the phonons
on the excitonic system during the presence of the pulse is negligible. This
is a result of the high-frequency cut-off in the phonon spectral density
[Eq.~(\ref{eq:phonon-spect-anal})], which leads to a decoupling at times
$t\lesssim \omega_c^{-1}$. Closed form results for an arbitrary sequence of
ultrashort excitations have been derived using a generating functions
formalism \cite{vagov2002ele,axt2005pho}. For preparation purposes, on the
one hand, a short preparation time is desired. However, ultrashort
excitations strongly restrict the available preparation schemes such that
essentially only the traditional Rabi flopping strategy, as discussed in
detail in Sec.~\ref{sec:resonant}, can be realized. Other schemes that turn
out to be advantageous for certain purposes (cf.~Secs.~\ref{sec:chirp} and
\ref{sec:detuned}) require longer pulse durations. For such excitations, one
has to rely on numerical methods. Most often, these methods fully account for
the dot-light coupling but treat the exciton-phonon interaction within
further approximations. Examples of such approaches are the
time-convolutionless approach
\cite{breuer2002the,machnikowski2008the,gawarecki2012dep}, the correlation
expansion
\cite{rossi2002the,forstner2003pho,hohenester2004qua,krugel2005the,krugel2006bac},
different types of master equations
\cite{breuer2002the,ramsay2010dam,mccutcheon2010qua,mccutcheon2011gen,roy2011pho,roy2012pol,harsij2012inf}
or time-dependent perturbation theory
\cite{machnikowski2004res,krugel2005the}. For the Hamiltonian dynamics within
the model established by Eqs.~(\ref{eq:Hel}), (\ref{eq:Hdip}) and
(\ref{eq:Hph}), also a numerically complete treatment without any further
approximations to the light-matter or the carrier-phonon coupling has been
implemented using a real-time path-integral approach
\cite{vagov2006hig,vagov2007non,vagov2011rea,thorwart2005non,glassl2012imp}.
Taking these numerically complete solutions as a benchmark, direct
comparisons have been performed for the correlation expansion
\cite{glassl2011lon,vagov2011dyn}, the time-convolutionless formulation
\cite{mccutcheon2011gen}, and master equation \cite{mccutcheon2011gen}
approaches. These comparisons revealed that for parameters typical for state
preparation protocols, i.e., sufficiently low temperatures, weak
carrier-phonon couplings and not too long times, the correlation expansion,
the time-convolutionless formulation as well as advanced master equation
approaches essentially give the same results. A direct comparison of the
correlation expansion and perturbation theory revealed that the latter fails
much earlier than the other approaches \cite{krugel2005the}, which is of
course not unexpected.

\begin{figure}[ht!]
\centering
\includegraphics[width=7cm]{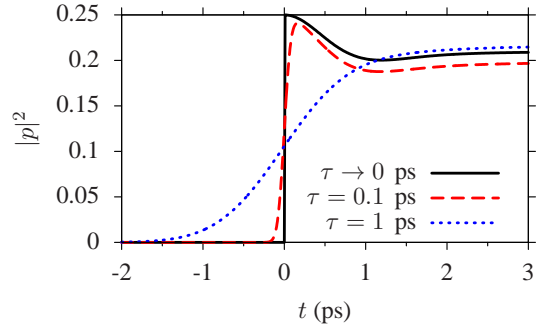}
\caption{\label{fig:polarization}
Modulus square of the polarization as a function of time
after resonant excitation of a QD
with an ultrashort pulse, a $100$~fs pulse and a $1$~ps pulse. All pulses have
a pulse area $\theta=\pi/2$.}
\end{figure}

A characteristic feature of the model defined by Eqs.~(\ref{eq:Hel}),
(\ref{eq:Hdip}), and (\ref{eq:Hph}) is the fact that in the basis of the
electronic states $|\nu\rangle$ the phonons only exhibit a diagonal coupling,
i.e., they do not give rise to transitions between different states. Thus,
after an optical excitation -- when the light pulse has gone -- the
occupations of the states remain constant. However, the optical polarization,
describing the coherence between the optically coupled states, exhibits an
initial decay on a picosecond time scale. This reflects the pure dephasing
nature of the model. As an example, the black solid line in
Fig.~\ref{fig:polarization} displays the time evolution of the modulus square
of the optical polarization of a two-level dot generated by an ultrafast
laser pulse with a pulse area $\pi/2$. Like in an ideal TLS, an initial value
of $0.25$ is reached followed by a decay which is not only non-exponential
but also only partial, i.e., the polarization approaches a
temperature-dependent finite value at long times. The origin of this behavior
as well as the dependence on the duration of the exciting pulse will be
discussed in more detail in Sec.~\ref{sec:superposition}. Obviously, such a
non-exponential and incomplete decay of the polarization, which in the
spectral domain corresponds to strongly non-Lorentzian line shapes of the
absorption or luminescence spectra
\cite{krummheuer2002the,besombes2001aco,forstner2002lig}, cannot be captured
by a simple decay rate for the polarization. In fact, a naive calculation of
phonon-induced relaxation rates according to Fermi's golden rule would give a
zero result in this case, because the energies of the initial and final
electronic state are the same and the phonon spectral density $J(\omega)$ is
zero at $\omega=0$. Thus, the polarization decay represents  a genuine
non-Markovian dynamics. We note in passing, that the non-exponential decay
towards a finite value is special for the superohmic coupling. In fact, in
the subohmic case an exponential decay is found even when $J(\omega)$ still
approaches zero for $\omega \to 0$ \cite{weiss2008qua}.

In view of these features, it may at first seem surprising, that even rather
simple Markovian master equation approaches can, under certain circumstances,
well reproduce experimental observations such as the damping of Rabi
oscillations. Here, it should be noted that provided the laser envelope is
slowly varying in time and the carrier-phonon coupling is not too strong, the
dressed states become stable quasiparticles. As the light-matter interaction
is already accounted for in the definition of the dressed states, the energy
of the incoming photons is included in the corresponding energies. Therefore,
the splitting between the dressed state energies is determined by the Rabi
frequency which, for near resonant excitation is typically of the order of a
few meV, i.e., of the order of typical acoustic phonon energies. Therefore,
the phonon coupling can now induce real transitions between the dressed
states which can be described by a transition rate. Consequently, for
moderate carrier phonon coupling strengths the phonons essentially lead to a
thermalization dynamics in the dressed state basis \cite{glassl2011lon} which
is often well represented by a Markovian rate equation. A more detailed
analysis based on master equations accounting for the phonon-induced memory
reveals that the simple Markovian treatment comes to its limit, e.g., when
phonon-induced renormalizations of the Rabi frequency become important: for
stronger carrier-phonon coupling the stable quasiparticles assume a polaronic
character, which can be used to formulate more advanced master equation
approaches that rely on polaron transformations
\cite{mccutcheon2010qua,mccutcheon2011gen,roy2011pho}.

\section{State preparation schemes}
\label{sec:schemes}

The state preparation of a QD typically aims at exciting the QD from its
ground state, i.e., when no exciton is present, into a specific final state,
e.g., a single exciton state, the biexciton state or a well-defined
superposition state. In this section we will discuss three basically
different optical control schemes and focus on the influence of the phonons
on the state preparation. The most common excitation scheme relies on the
coherent manipulation of the few-level system by resonant laser pulses. The
pulses induce Rabi oscillations in the carrier system and drive the QD into a
final state, which strongly depends on the pulse and material parameters such
as pulse shape, pulse duration and dipole coupling matrix elements. The
second scheme uses chirped pulses where the frequency sweeps over the
resonance frequency of the QD starting either below or above the resonance.
Here the excitation mechanism is an adiabatic rapid passage (ARP) from the
ground state to the exciton or biexciton state which, as soon as the
adiabaticity condition is reached, only weakly depends on the exact pulse
parameters like the pulse area or the chirp rate. Both these schemes have in
common that they work perfectly in the case of an isolated few-level system.
Phonons deteriorate the controllability and therefore the study of the
influence of phonons is mainly directed towards finding parameter ranges
where this influence is of minor importance. In contrast to these two
coherent schemes, the third excitation scheme is based on incoherent
phonon-induced transitions to prepare the exciton or biexciton state by an
excitation with a detuned laser pulse. This scheme therefore relies on the
carrier-phonon interaction and works even better for stronger couplings. In
the present section we will describe these schemes and summarize the
experimental and theoretical work that has been performed in these fields.

\subsection{Resonant excitation}
\label{sec:resonant}

When a QD modeled as an ideal TLS is excited resonantly, i.e., by
monochromatic light with the frequency $\omega_L=\omega_X$, the occupations
of the two states, here the ground and the exciton state, oscillate between
zero and one as a function of time. These are the well-known \emph{Rabi
oscillations}, which were first described for spin oscillations in a rotating
magnetic field \cite{rabi1937spa}. Without any further couplings the
oscillation continues without any damping as long as the light field is
present. A different but completely equivalent picture of Rabi oscillations
is provided in the framework of the dressed state basis introduced in
Sec.~\ref{sec:model}. Here the exciting light field gives rise to the
splitting of the dressed states by the Rabi frequency and Rabi oscillations
correspond to a quantum beat of these two states. The coupling to phonons
disturbs these ideal Rabi oscillations. It is instructive to first discuss
the influence of the phonons on Rabi oscillations in the time regime to
understand their influence on the state preparation.

\subsubsection{Rabi oscillations}

\begin{figure}[!ht]
\centering
\includegraphics[width=7cm]{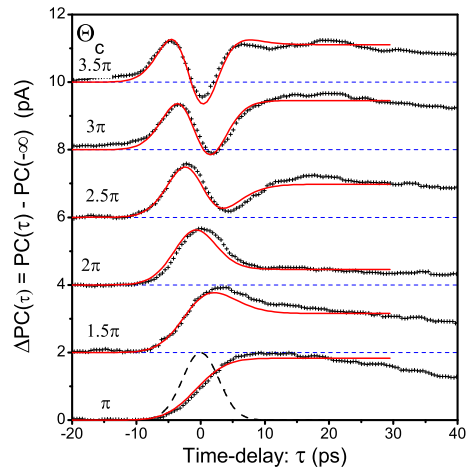}
\caption{\label{fig:rabiexp}
Experimentally measured Rabi oscillations as a function of time for increasing laser power.
The dashed line marks the envelope of the laser pulse. The measurement has been
performed by photocurrent detection of the time-dependent Autler-Townes-splitting
from the exciton-to-biexciton transition. Figure reprinted with permission from Boyle
\emph{et al} \cite{boyle2009bea}.}
\end{figure}

The measurement of time-resolved Rabi oscillations for a single QD is
experimentally challenging. To measure the time dependence of the Rabi
oscillations, the change in the Autler-Townes-splitting in the TLS formed by
the exciton-biexciton transition has been detected \cite{boyle2009bea}. The
results are shown in Fig.~\ref{fig:rabiexp}. The Autler-Townes splitting
refers to the splitting of the transition line caused by the coupling to the
light field, when observed by a third (auxiliary) state \cite{ramsay2010are}.
The measurement has been performed using photocurrent detection
\cite{boyle2008two} with an excitation with a finite pulse length as
indicated in the lower part of Fig.~\ref{fig:rabiexp}. When the laser power
is increased, the Rabi frequency increases and more Rabi oscillations can
take place during the pulse. Furthermore, the Rabi oscillations are damped
and the damping increases with increasing power. Time-resolved Rabi
oscillations have also been observed experimentally for a QD placed in a
microcavity where the light-matter interaction was in the strong coupling
regime. Here, the Rabi oscillations reveal themselves in the time-resolved
correlation function \cite{muller2007res,xu2007coh,flagg2009res}. Recently
time-resolved Rabi oscillations have been measured by monitoring the
time-dependent resonance fluorescence from a singly charged QD, where a TLS
is formed by the negatively charged ground state and the trion state
\cite{schaibley2013dir}.

The coupling to phonons has several effects on the Rabi oscillations. First,
it leads to a damping of the oscillations. The origin of this damping is most
easily understood in the dressed state picture. In the case of Rabi
oscillations the electronic system is in a superposition of the two dressed
states. Phonons give rise to incoherent transitions between the dressed
states. At low temperatures, when essentially only phonon emission processes
are present, this leads to a relaxation to the lower dressed state, while at
higher temperatures, when emission and absorption processes are almost
balanced, both dressed states will be equally populated \cite{glassl2011lon}.
In both cases the coherence between the states, which is responsible for the
oscillations, decays.

In addition to the damping there are other phonon-induced phenomena. Due to
the coupling to phonons, the excitation of the QD leads to a local lattice
distortion in the region of the QD caused by the change of the electronic
charge distribution. The resulting exciton-phonon complex is known as the
\emph{polaron}. The buildup of the polaron leads to a shift of the transition
line from the bare value $\omega_X$ to the polaron-shifted value $\omega_X -
\sum_{\mathbf{q}} \frac{|g_{\mathbf{q}}|^2 }{\omega_{\mathbf{q}} }$
\cite{krummheuer2002the}. Accordingly, when modeling resonant excitations
based on the Hamiltonian in Eqs.~(\ref{eq:Hel}),(\ref{eq:Hdip}) and
(\ref{eq:Hph}), this shift has to be taken into account by adjusting the
light frequency $\omega_L$ to the polaron-shifted transition frequency. The
shift of the exciton line caused by the exciton-phonon interaction can be
directly seen in experiments measuring either the passage of a strain pulse
through the dot \cite{akimov2006ult,gotoh2013mod} or by applying a periodic
strain modulation in terms of a surface acoustic wave, where it leads to
sidebands in the QD fluorescence spectrum \cite{metcalfe2010res}. It can be
even exploited for an ultrafast switching of a QD placed in a microcavity
into the lasing regime by shifting the transition frequency into resonance
with the cavity mode \cite{bruggemann2011las}.

Finally, the polaronic nature does not only modify the energies of the states
but also the coupling to the light field. This renormalization of the
exciton-light interaction strength gives rise to a change in the frequency of
the Rabi oscillations compared to its phonon-free value. This leads in
particular to a temperature dependence of the Rabi frequency even if the
light field is kept constant \cite{machnikowski2004res,krugel2005the}.

\begin{figure}[!ht]
\centering
\includegraphics[width=7cm]{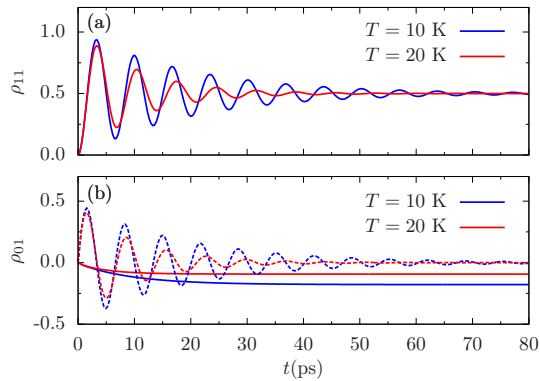}
\caption{\label{fig:rabiosc}
Dynamics of the TLS coupled to phonons for the case of resonant
excitation by a constant light field switched on at $t=0$.
(a) Occupation of the exciton state $\rho_{11}$ and (b)
real (solid) and imaginary (dashed) part of the coherence $\rho_{01}$ in a
frame rotating with the light field. Blue and
red curves correspond to the temperatures $10$~K and $20$~K, respectively.
Figure reprinted with permission from Gl{\"a}ssl \emph{et al} \cite{glassl2011lon}.}
\end{figure}

An example of the dynamics of the two-level QD coupled to phonons is shown in
Fig.~\ref{fig:rabiosc}, where the exciton occupation and the interband
polarization, i.e., the coherence between ground and exciton state are
displayed for the case of driving by a resonant light field with constant
amplitude. Note that the polarization is displayed in a frame rotating with
the light field. The Figure clearly shows the damping of the oscillations and
the approach of a stationary state on a time scale of several tens of
picoseconds. Both the damping rate and the oscillation frequency are
temperature dependent, as discussed above. In the stationary state the
occupation of the exciton level is $0.5$ independent of the temperature, and
the imaginary part of the polarization vanishes. The stationary value of the
real part of the polarization, on the other hand, depends on the temperature.
It has been found that the stationary state can be very well understood as a
thermal distribution over the dressed states, confirming again the usefulness
of this picture. Only for very low temperatures slight deviations have been
found \cite{glassl2011lon}. Here, both path integral and correlation
expansion calculations revealed a stationary value of the polarization lying
between the predictions by a weak coupling theory \cite{weiss2008qua} and the
thermal distribution over the dressed states.

For light, which is not strictly circularly polarized, in general both
exciton states and also the biexciton are excited. Typically the
exciton-to-biexciton transition energy lies below the ground state-to-exciton
transition energy by a few meV, the biexciton binding energy $\Delta_B$
\cite{borri2002rab,boyle2008two,zecherle2010ult}, such that a two-photon
process can be chosen which is resonant on the ground state-to-biexciton
transition but where the single-photon frequency is sufficiently detuned from
the ground state-to-exciton and exciton-to-biexciton frequencies. This leads
to two-photon Rabi oscillations between the ground and biexciton state
\cite{stufler2006two,bensky2013hig,muller2014ond,sun2014pha}. The impact of
phonons is similar as in the two-level case \cite{machnikowski2008the}. The
stationary values of the occupation and also the time scale on which the
stationary state is reached, depend strongly on the polarization of the
exciting laser pulses \cite{glassl2012imp, glassl2012pol}.

\subsubsection{Rabi rotations}

\begin{figure}[!ht]
\centering
\includegraphics[width=7cm]{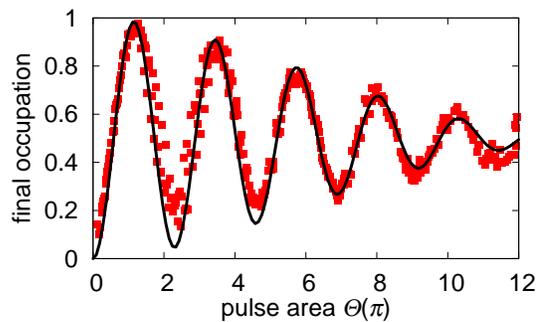}
\caption{\label{fig:rabirot}
Occupation of the exciton state as a function of the pulse area at $T=5$~K
obtained from photocurrent measurements
(dots) and calculated using the model discussed in the paper (solid line).
We thank A.~J.~Ramsay for providing us with the
experimental data originally published in \cite{ramsay2010pho}.}
\end{figure}

For an excitation of the QD with laser pulses the dynamics of the occupations
are restricted to the time when the light field couples to the QD. After the
pulse the occupations remain constant as long as radiative decay can be
disregarded. Therefore it is meaningful to look at the final occupation after
such an excitation. This is indeed the typical type of experiment to monitor
the Rabi-type dynamics of excitons and biexcitons. Typically the final
occupation is plotted as a function of the pulse area, which for a fixed
pulse duration is proportional to the square root of the light intensity.
Also here an oscillatory behavior shows up, which is commonly named
\emph{Rabi rotation} because in a TLS it reflects the rotation of the Bloch
vector into a new direction by the action of the pulse. An example of both
measured and calculated Rabi rotations is shown in Fig.~\ref{fig:rabirot}. In
an ideal TLS pulse areas given by odd multiples of $\pi$ result in a complete
excitation of the system (occupation of one) while for even multiples of
$\pi$ the system is completely driven back into the ground state. Rabi
rotations are a widely used tool for state preparation, because they allow a
preparation of pure exciton or biexciton states as well as of coherent
superpositions with arbitrary amplitudes of ground, exciton and biexciton
state to be carried out. The price which has to be paid for this versatility
is the very sensitive dependence of the dynamics and thus of the generated
state on the pulse parameters such as pulse shape, intensity and duration, as
well as on material parameters like the energies or dipole coupling matrix
elements. Already small deviations from the resonance or from the correct
pulse area have a large influence on the state preparation. Therefore, for an
ensemble of QDs, which typically have a range of transition frequencies and
dipole matrix elements, a controlled state preparation using Rabi rotations
is not possible.

First experiments showing Rabi rotations in a single QD, being the prototype
of a coherent manipulation of a QD, have been realized more than ten years
ago in different structures
\cite{stievater2001rab,kamada2001exc,htoon2002int,zrenner2002coh,besombes2003coh}.
Since then, many other experiments have been performed and nowadays Rabi
rotations are commonly used in many optical control experiments
\cite{stufler2005qua,wang2005dec,kuroda2007fin,melet2008res,takagi2008coh,ravaro2010dec,ramsay2010pho,ramsay2010dam,zecherle2010ult,wolpert2012ult,monniello2013exc}.

The coupling of the QD to its environment leads to a temporal damping of Rabi
oscillations, as discussed above. This damping is reflected in a
corresponding damping of Rabi rotations when plotted as a function of the
pulse area, which has been investigated in a variety of experimental studies.
Depending on the sample structure and the excitation conditions different
coupling mechanisms may be responsible for the observed damping. For example,
the intensity-dependent damping of Rabi rotations with a pulse area up to
$10\pi$ has been attributed to simultaneous bound-to-continuum transitions
involving wetting layer states \cite{wang2005dec}. A very good agreement of
the damping behavior of measured Rabi rotations with pulse areas up to
$12\pi$ and theoretical calculations including the electron-phonon coupling
as described in Sec.~\ref{sec:model} has been recently demonstrated
\cite{ramsay2010dam,ramsay2010pho} (see also Fig.~\ref{fig:rabirot}). The
calculations were able to reproduce the experimental data for a large
temperature range from 5~K up to 75~K. Also for QDs embedded in a
one-dimensional wave guide the main source for the damping of Rabi rotations
has been attributed to the coupling to acoustic phonons with additional
contributions resulting from the enhanced radiative decay due to the strong
coupling between the QD and the optical wave guide mode
\cite{monniello2013exc}.

The experimental realization of Rabi rotations between the biexciton state
and an exciton state \cite{li2003ana}, as well as between the biexciton and
the ground state
\cite{stufler2006two,bensky2013hig,muller2014ond,sun2014pha,flissikowski2004two,jayakumar2013det},
which can be both described in the four-level model, has great potential
impact on the field of quantum logic where exciton-biexciton systems have
been proposed as two-qubit quantum gates \cite{troiani2000exp,li2003ana} or
as emitters of entangled photon pairs
\cite{benson2000reg,moreau2001qua,santori2002pol,akopian2006ent,stevenson2006ase,hafenbrak2007tri,dousse2010ult,muller2014ond}.
Rabi rotations between the ground and the biexciton state were also preformed
by using two-color excitations \cite{boyle2010two}. A recent review of
coherent optical control of QDs can be found in Ref.~\cite{ramsay2010are}.

The impact of the interaction of a QD exciton with acoustic phonons on the
damping of Rabi rotations has been the subject of many theoretical works
\cite{forstner2003pho,machnikowski2004res,wu2004dyn,krugel2005the,axt2005pho,machnikowski2008the,mccutcheon2010qua,mccutcheon2011gen,vagov2011dyn,glassl2011inf,vagov2007non,reiter2012pho}.
Besides analyzing specific experimental conditions, these studies have also
been motivated by the exemplary character of the Hamiltonian describing a
bosonic bath coupled to a two-level system
\cite{weiss2008qua,mahan2000man,breuer2002the}. More general types of
coupling to a bosonic bath have also been investigated
\cite{mogilevtsev2008dri}, however, here we will restrict ourselves to the
type of coupling introduced in Sec.~\ref{sec:model}.

\begin{figure}[!ht]
\centering
\includegraphics[width=7cm]{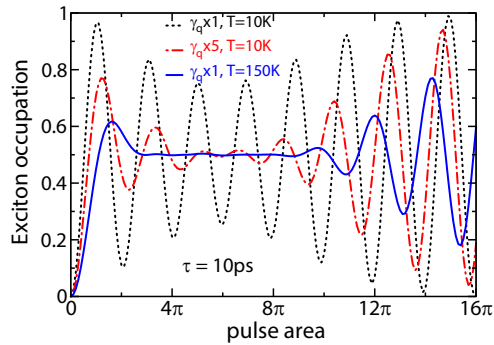}
\caption{\label{fig:reapp} Exciton occupation as a function of the
pulse area in the TLS calculated for the case of excitation by
a $10$~ps rectangular pulse at different temperatures and coupling strengths.
Here, $\gamma_{\mathrm{q}}$ refers to the coupling $|g_{\mathrm{q}}|^2$ with the
standard material GaAs parameters. For the dashed-dotted line this value
has been increased by a factor of 5. Figure reprinted with
permission from Vagov \emph{et al} \cite{vagov2007non}.}
\end{figure}

Like for Rabi oscillations in the time domain, a renormalization of the
frequency, here with respect to the pulse area, is found. However, while Rabi
oscillations are always decaying in time, for Rabi rotations the damping has
been found to be a non-monotonic function of the pulse area. For small pulse
areas the amplitude of the Rabi rotations decreases with increasing pulse
area thus showing the expected damping behavior, while for high pulse areas
the amplitude increases again, which resembles an undamping. However, it
should be noted that this is not an undamping in the time regime such as,
e.g., the revival of Rabi oscillations in the Jaynes-Cummings model
\cite{gerry2005int}. This behavior has therefore been termed the
\emph{reappearance} of Rabi rotations \cite{vagov2007non}; an example is
shown in Fig.~\ref{fig:reapp}.

The non-monotonic behavior of the damping can be traced back to the resonant
nature of the exciton-phonon coupling matrix elements as reflected in the
phonon spectral density given by Eq.~(\ref{eq:phonon-spect}) (see also
Fig.~\ref{fig:spectraldensity}). The spectral density clearly reveals that
only phonons in a certain frequency window are efficiently coupled to the
exciton and that there is a frequency $\omega_{\mathrm{max}}$ where the
coupling is most efficient. As seen in Fig.~\ref{fig:spectraldensity}, this
frequency depends on the QD size; it can be roughly identified as the
frequency of phonons with a wave length of the order of the QD size. The
non-monotonic behavior of the damping can now be understood in terms of a
resonance between the exciton dynamics characterized by the Rabi frequency
$\Omega_{\mathrm{Rabi}}$ and the dynamics of the most strongly coupled
phonons characterized by $\omega_{\mathrm{max}}$
\cite{machnikowski2004res,krugel2005the,wigger2014ene}. If these two
frequencies coincide, the damping will be strongest. For much smaller Rabi
frequencies ($\Omega_{\mathrm{Rabi}} \ll \omega_{\mathrm{max}}$) the
electronic system oscillates so slowly that the phonon system follows
essentially adiabatically and no energy is transferred irreversibly to the
phonons. On the other hand, for Rabi frequencies much higher than the
resonance ($\Omega_{\mathrm{Rabi}} \gg \omega_{\mathrm{max}}$) the change in
the electronic system is so fast that the phonons cannot follow leading again
to an inefficient coupling.

Instead of using the resonance argument in the time domain, also the spectral
domain can be employed for the interpretation of the non-monotonic dependence
of the damping on the Rabi frequency \cite{krugel2005the}. The driving light
field gives rise to the appearance of dressed states with an energy splitting
given by the Rabi frequency $\Omega_{\mathrm{Rabi}}$. Phonon-induced
transitions between the dressed states occur with a rate which is
proportional to the phonon spectral density at the frequency corresponding to
this energy splitting, which again explains why these transitions, and
therefore the damping, are most pronounced when the Rabi frequency coincides
with the position of the maximum of the spectral density, i.e., when
$\Omega_{\mathrm{Rabi}} \approx \omega_{\mathrm{max}}$.

The non-monotonic damping and the reappearance of Rabi rotations are a
general behavior related to the structure of the exciton-phonon coupling in
QDs. As is seen in Fig.~\ref{fig:reapp}, it also occurs for a coupling which
is much stronger than the one in typical QD structures and at rather high
temperatures, when there are essentially no more oscillations visible in a
certain window of Rabi frequencies. The details of the occurrence of the
reappearance, however, strongly depend on many details of the exciting laser
pulse, e.g., the pulse duration or the pulse shape. For pulses shorter than
$\omega_{\mathrm{max}}^{-1}$ already for a $\pi$-pulse the excitonic dynamics
is faster than the phonon dynamics such that already the first rotation is in
the regime of decreasing damping. Therefore, Rabi rotations for short pulse
excitations are essentially undamped \cite{krugel2005the}. The pulse shape
determines the distribution of Rabi frequencies during the action of the
pulse. A rectangular pulse is characterized by a single, fixed Rabi frequency
during the pulse. Therefore there is a clear resonance and the reappearance
sets in immediately above the resonance. In a Gaussian pulse, on the other
hand, the Rabi frequency increases continuously from zero to a maximal value
and then decreases again. Therefore, even if the amplitude at the pulse
maximum is already beyond the resonance there are parts of the pulse in the
leading and trailing edge where the amplitude satisfies the resonance
condition and therefore a pronounced damping is present during these times.
For this reason, in the case of a Gaussian pulse the amplitude of the Rabi
rotations grows much slower after the onset of the reappearance and much
higher pulse areas are required to reenter the regime of pronounced Rabi
rotations \cite{glassl2011inf}.

Recent experiments on Rabi rotations with pulse areas up to $12\pi$ have been
quantitatively described by the model on which the present review is based
\cite{ramsay2010pho,ramsay2010dam}. Indeed for small pulse areas the
influence of phonons is small and increases with increasing pulse area. The
regime of reappearance of Rabi rotations has not yet been clearly reached in
these experiments, even though a detailed comparison with calculations
suggests that some features of the measurements can be regarded as first
hints for an onset of the reappearance.

Both experimental and theoretical studies thus show that in order to achieve
a high-fidelity state preparation based on a resonant excitation it is
necessary to carefully select the excitation conditions regarding pulse
length, pulse shape and pulse area. The optical control typically performs
better when choosing short, rectangular pulses and working within a pulse
area range up to $2\pi$. Of course, when optimizing the excitation conditions
the limitations of the model have to be kept in mind. For long pulses, other
dephasing or relaxation mechanisms such as radiative decay may become
effective \cite{alicki2004opt}. Short pulses with their broad spectrum may
excite other transitions either in the same dot or in nearby dots which also
deteriorates the preparation of the desired state.

Currently also the optical control of QDs in a microcavity has come into the
focus of attention, where vacuum Rabi oscillations and the corresponding
vacuum Rabi splitting are seen as indications for reaching the strong
coupling regime between light and matter
\cite{reithmaier2004str,yoshie2004vac,peter2004exc,hennessy2007qua}. In the
case of strong light-matter coupling there is a strong interplay between
exciton-phonon and light-matter interaction, which for example can be seen in
the line-width broadening in the Mollow-Triplet
\cite{ulhaq2010lin,ulrich2011dep}. The combined exciton-photon and
exciton-phonon interaction in the strong coupling regime has been subject to
many theoretical works
\cite{zhu2003eff,carmele2010ant,kaer2010non,kaer2012mic,kaer2013mic,roy2011pho,roy2011inf,roy2012pol,roy2012ano,glassl2012int}.
A detailed discussion of the phenomena related to the interplay between these
coupling mechanisms is beyond the scope of this review.

\subsubsection{Generation of superposition states}
\label{sec:superposition}

So far we have mostly discussed the influence of phonons on the occupation of
the QD states. For applications of optically controlled QDs in quantum
information processing it is also necessary to prepare a given superposition
state, e.g., a superposition of ground and exciton state
\cite{bonadeo1998coh}. One way to create a superposition state is to use a
Rabi rotation with a pulse area, which is not equal to a multiple of $\pi$.
For example, a $\pi/2$ pulse applied to a TLS creates a superposition between
ground and exciton state $(|G\rangle + i |X\rangle)/\sqrt{2}$. In the
language of quantum information processing applying a $\pi/2$ pulse
corresponds to the action of a Hadamard gate on a qubit consisting of the
states $|G\rangle$ and $|X\rangle$ \cite{nielsen2000qua}.

The ability to fully control the TLS is often demonstrated in Ramsey
interference experiments, where a pair of two $\pi/2$ pulses is used to
resonantly excite the QD
\cite{htoon2002int,stufler2005qua,stufler2006ram,michaelis2010coh,ramsay2010are}.
If the second pulse is in phase with the polarization created by the first
pulse, there is constructive interference driving the system into the exciton
state $|X\rangle$. If the second pulse has the opposite phase as the
polarization there is destructive interference; the exciton is removed from
the QD and the ground state $|G\rangle$ is restored. When plotted as a
function of the phase difference between the pulses this leads in an ideal
TLS to oscillations of the exciton occupation between zero and one. In the
presence of dephasing processes these oscillations are damped.

Also for the preparation of superposition states the phonons are a major
source of dephasing. Besides the dephasing during the action of the pulse,
which can be reduced by using short pulses, the polarization is in general
subject to dephasing even after the preparation of the state. This becomes
clear for an excitation with ultrafast laser pulses, where the phonons cannot
act during the pulse. After the generation of a superposition state the
occupations remain constant, but the polarization decays as shown in
Fig.~\ref{fig:polarization} for the case of an excitation by pulses with a
pulse area of $\pi/2$ and three different pulse durations. In the case of
ultrafast excitation the maximal polarization of $0.25$ is reached. Directly
after the generation a fast initial decay of the polarization within a time
scale of about $1$~ps is found \cite{krummheuer2002the}. In the case of the
$100$~fs pulse a slightly lower value is reached, but the behavior is still
similar to the ultrafast limit. Because the ultrafast excitation of an
exciton leads to an almost simultaneous change of the charge density, the
equilibrium position of the crystal lattice changes. Due to the abrupt
change, coherent phonons are created and in the case of acoustic phonons a
coherent wave packet is sent out of the dot which carries away information
from the QD
\cite{machnikowski2004res,roszak2006whi,wigger2013flu,wigger2014ene}. This
leads to the observed dephasing of the coherence
\cite{krummheuer2002the,krummheuer2005pur}. After the wave packet has left
the dot, the polarization exhibits no further decay caused by the phonons.
The time scale of the initial decay is determined by the time the wave packet
needs to leave the QD. Other interaction mechanisms, in particular radiative
decay, give rise to an exponential decay on a much longer time scale
\cite{borri2001ult}. The non-exponential decay of the polarization, which is
a clear indication of the non-Markovian nature of the exciton-phonon coupling
in the absence of a driving light field, has been measured using
four-wave-mixing techniques
\cite{borri2001ult,borri2007fou,patton2006tim,kasprzak2011coh,kasprzak2013vec}.

From the polarization dynamics after an ultrafast excitation the absorption
spectrum of the QD can be obtained, which consists of the zero phonon line
and the typical phonon background resulting from the interaction with
acoustic phonons. The interaction with optical phonons gives rise to
satellites in the spectrum separated by multiples of the optical phonon
frequency, which again are superimposed on a background due to the acoustic
phonons. This type of spectra has been seen in many experimental and
theoretical studies
\cite{besombes2001aco,borri2001ult,krummheuer2002the,stock2011aco,ostapenko2012exc}.

For longer pulses, the change of the electronic charge density caused by the
excitation of the QD is much slower and, for sufficiently long pulses in the
adiabatic limit no wave packet is emitted. Thus, in the polarization no decay
of the polarization after the pulse is found, as shown in
Fig.~\ref{fig:polarization} for the case of an excitation by a $1$~ps pulse.
Instead, coherence is lost already during the action of the pulse, which also
results in a reduced occupation of the states. Interestingly, the final
polarization long after the pulse, say for $t>2$~ps, is larger in the case of
the longer pulse. Instead of using a single long pulse a sequence of
ultrashort pulses can be used, which for suitable parameters also gives rise
to a reduced decoherence compared to a single ultrashort pulse
\cite{axt2005red,hodgson2008dec}.

\begin{figure}[!ht]
\centering
\includegraphics[width=7cm]{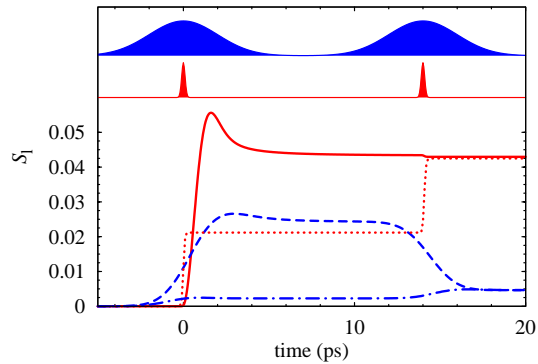}
\caption{\label{fig:dressed_qubit} Linear entropy as a function of time for
the case of excitation by two $\pi/2$ pulses. Red (solid and dotted) curves:
excitation by two $100$~fs pulses, blue (dashed and dashed-dotted) curves:
excitation by two $2$~ps pulses. The solid and dashed lines refer to a calculation
of the entropy for the bare qubit (ground and exciton state), the dotted and
dashed-dotted lines are calculated for the dressed qubit (ground and polaron
state). The pulse envelopes are shown in the upper part (in arbitrary units).
Figure reprinted with
permission from Machnikowski \emph{et al} \cite{machnikowski2007qua}.}
\end{figure}

It is interesting to note that the reduced value of the polarization after
the longer $\pi/2$ pulse seen in Fig.~\ref{fig:polarization} compared to the
ideal value of $0.25$ is not completely related to irreversibility, in
contrast to the decay after an ultrafast excitation, where irreversibility
results from the emission of the phonon wave packet. This is seen when
comparing a Ramsey-type interference using short and long pulses
\cite{machnikowski2007qua} and calculating the linear entropy. The linear
entropy $S_1=1-\ln(\rho^2)$, where $\rho$ is the reduced density matrix in
the QD subspace, is a suitable measure for the purity of a QD state
\cite{breuer2002the}. The results of such calculations are shown in
Fig.~\ref{fig:dressed_qubit} (solid and dashed lines). The first pulse drives
the system into an entangled state between exciton and phonons and thus
increases the linear entropy. While in the case of short pulses the linear
entropy essentially remains constant after the second pulse, for long pulses
it is considerably reduced by the second pulse. In fact, it turns out that in
the case of an excitation by long pulses the appropriate definition of a
qubit should include the polaron dressing of the exciton
\cite{machnikowski2007qua}. The dotted and dashed-dotted lines in
Fig.~\ref{fig:dressed_qubit} show the results for the linear entropy when the
reduced density matrix is calculated in the polaron basis. In this case, the
entropy after the first pulse is much smaller than in the bare exciton basis
(solid and dashed lines), in particular for the long pulse, and the entropy
never decreases. The dressed qubit therefore undergoes much lower dephasing
and much higher fidelities of quantum gate operations can be achieved.
However, when strongly increasing the pulse duration other dephasing
mechanisms such as radiative decay usually become more important. Taking into
account this trade-off between different types of decoherence then leads to
optimal pulse durations which depend on the specific QD structure as well as
on temperature \cite{alicki2004opt}.

\subsection{Excitation with chirped laser pulses}
\label{sec:chirp}

A different method of state preparation is achieved by using \emph{chirped}
laser pulses. During these frequency-modulated pulses the frequency varies
with time. In the most simple case the modulation is linear, such that for a
pulse with pulse maximum at time $t=0$ the corresponding laser frequency is
given by $\omega(t) = \omega_0 + at$, where $\omega_0$ is the central
frequency of the laser pulse at the pulse maximum, and $a$ is the chirp rate.
In most cases, $\omega_0$ is taken to be resonant with the exciton
transition. For the generation of chirped laser pulses a transform-limited
pulse is sent through a chirp filter characterized by a chirp coefficient
$\alpha$ \cite{malinovsky2001gen,melinger1994adi}. In such a filter an
initial Gaussian pulse with amplitude $f_0$ and pulse duration $\tau_0$ is
transformed into a chirped Gaussian pulse with pulse duration $\tau$, chirp
rate $a$ and amplitude $\tilde{f}$, given by
\begin{equation}
\tau=\tau_0\sqrt{1 + \frac{\alpha^2}{\tau_0^4}}, \quad
a=\frac{\alpha}{\alpha^2+\tau_0^4}, \quad \mathrm{and} \quad \tilde{f}=f_0 \sqrt{\frac{\tau_0}{\tau}},
\end{equation}
respectively \cite{saleh2007fun}.

\begin{figure}[!ht]
\centering
\includegraphics[width=7cm]{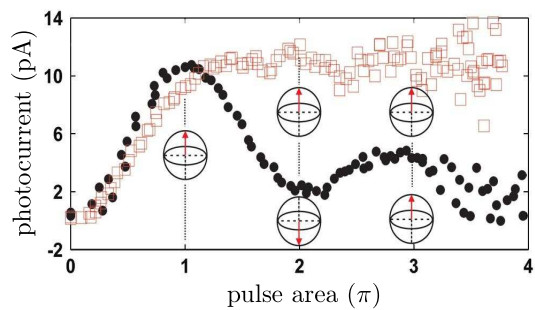}
\caption{\label{fig:arp_wu} Exciton occupation as a function of the pulse area
obtained from photocurrent measurements for the case of resonant excitation
by a $2$~ps transform-limited pulse (solid circles)
and an excitation by a $15$~ps chirped pulse  (open squares).
Figure reprinted with permission from Wu \emph{et al} \cite{wu2011pop}.}
\end{figure}

\begin{figure}[!ht]
\centering
\includegraphics[width=7cm]{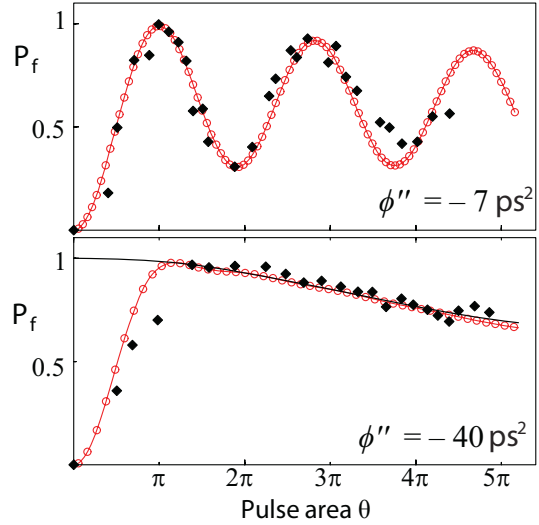}
\caption{\label{fig:arp_debnath} Exciton occupation $P_f$ as a function of the
pulse area for a very small (upper panel) and a large negative chirp (lower panel).
Diamonds show the experimental results from Ref.~\cite{simon2011rob}, circles
are obtained from numerical simulations and the solid lines shows results from
an analytical approximation valid in the adiabatic regime. Note that $\phi^{\prime\prime}$
is twice the chirp coefficient $\alpha$ defined in the text. Figure reprinted with
permission from Debnath \emph{et al} \cite{debnath2012chi}.}
\end{figure}

\begin{figure}[!ht]
\centering
\includegraphics[width=11cm]{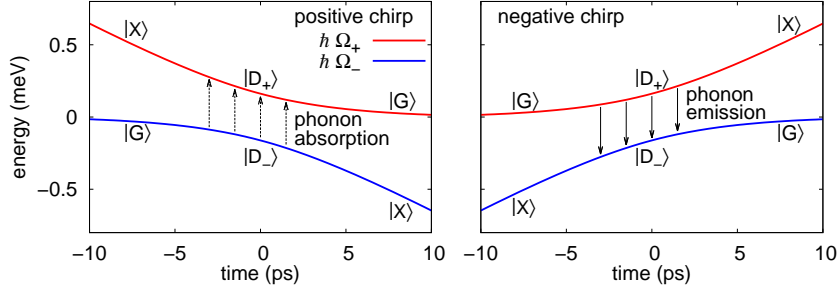}
\caption{\label{fig:dressed} Schematic plot showing the time evolution of the dressed states in a
two-level system excited by a chirped laser pulse for positive (left panel)
and negative (right panel) chirp.}
\end{figure}

The use of chirped laser pulses in atomic systems is a well established tool
\cite{vitanov2001laser,tannor2007int} and recently this technique has been
transferred to single QDs, where a stable population inversion has been
experimentally demonstrated using a linear chirp
\cite{wu2011pop,simon2011rob} (see Fig.~\ref{fig:arp_wu} and the experimental
points in Fig.~\ref{fig:arp_debnath}). The idea of population transfer via
adiabatic rapid passage (ARP) in a TLS can be easily understood in the
dressed state picture, as already introduced in Sec.~\ref{sec:model}. When
the system evolves adiabatically, at each time step a diagonalization of the
Hamiltonian can be performed leading to time-dependent dressed states. The
dressed states for a two-level QD under excitation with a linearly chirped
laser pulse with a duration of $5$~ps are shown in Fig.~\ref{fig:dressed}.
Before and after the pulse the dressed states can be identified with the
system eigenstates, namely the ground state $|G\rangle$ and the exciton state
$|X\rangle$. At the pulse maximum ($t=0$) both dressed states are given by
equal superpositions of $|G\rangle$ and $|X\rangle$. When following one of
the branches, it changes its character continuously from the ground state
$|G\rangle$ to the exciton state $|X\rangle$ or vice versa. The requirement
for ARP, namely that the system follows the branch adiabatically, is
quantified in the Landau-Zener criterion $|f|^2 \gg |a|$
\cite{tannor2007int,zener1932non,landau1932ath} and the condition $a\tau^2
\gg 1$ \cite{malinovsky2001gen}. Therefore, as soon as the adiabatic
threshold is overcome, by following one of the branches a robust and complete
population inversion of the system can be achieved. The robustness of ARP
refers to the fact, that variations in the pulse area, the pulse duration
and/or the chirp rate do not affect the state preparation, as long as the
adiabaticity condition is not violated. Because of this insensitivity ARP has
also been predicted for ensembles of QDs \cite{schmidgall2010pop}, provided
the distribution of energies and dipole matrix elements of the QDs is such
that the adiabaticity condition is fulfilled for the whole ensemble.

In an ideal TLS by using the ARP scheme, like in the case of resonant
excitation, an exciton preparation with arbitrarily high fidelity can be
achieved. However, also here phonons act as a limiting factor. Let us discuss
the influence of the phonons on the final occupation of the system after the
action of the laser pulse as a function of the pulse area. The experimental
data in the lower part of Fig.~\ref{fig:arp_debnath} show that the exciton
state occupation approaches unity for pulse areas between $\pi$ and $2\pi$
and then decreases with increasing pulse area. Such a decrease is absent in
an ideal TLS and is attributed to phonons.

\begin{figure}[!ht]
\centering
\includegraphics[width=7cm]{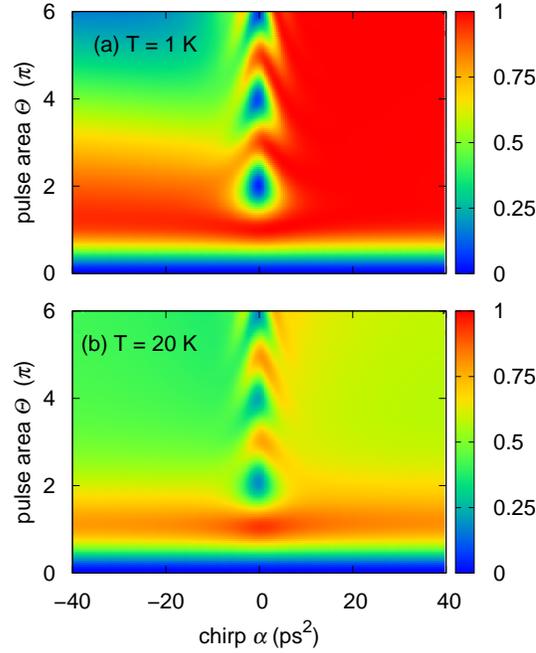}
\caption{\label{fig:arp_theo} Calculated exciton occupation as a function of pulse area
and chirp coefficient at two different temperatures. Figure reprinted with permission
from L{\"u}ker \emph{et al} \cite{luker2012inf}.}
\end{figure}

Theoretical results for the final exciton occupation at two different
temperatures and over a wide range of chirp coefficients and pulse areas are
shown in Fig.~\ref{fig:arp_theo}. These calculations reveal, that for low
temperatures and positive chirps the state preparation is almost not affected
by phonons, while for negative chirps a high-fidelity exciton generation is
restricted to pulse areas between about $\pi$ and $2\pi$. For higher pulse
areas the efficiency considerably drops even at low temperatures. This is in
agreement with the experimental and theoretical results shown in
Fig.~\ref{fig:arp_debnath}. The influence of the phonons on the ARP process
can be well understood in the dressed state picture. Here the phonons give
rise to transitions between the adiabatic eigenstates. A transition from the
upper to the lower branch occurs via the emission of a phonon, while the
reverse process is associated with the absorption of a phonon. At low
temperatures, when essentially no phonons are present, phonon absorption is
highly unlikely resulting in a strong asymmetry between the paths. In the
case of a positive chirp the system evolves along the lower branch and is
therefore almost unaffected by phonons, while for a negative chirp the
evolution is along the upper path and even at low temperatures phonon
emission leads to pronounced transitions to the lower branch. This asymmetry
has been found in all calculations based on different levels of complexity
\cite{luker2012inf,reiter2012pho,debnath2012chi,eastham2013lin} and has been
seen in recent experiments \cite{mathew2014sub}. The interpretation is
confirmed by an analytical model, where the phonon transition rates between
the dressed states are calculated from the phonon spectral density
[Eq.~(\ref{eq:phonon-spect})] at the frequency corresponding to the
instantaneous energy splitting of the dressed states \cite{debnath2012chi}.
The results of this model are shown as the solid black line in the bottom
panel of Fig.~\ref{fig:arp_debnath}. As soon as the condition for adiabatic
evolution is reached, i.e., for pulse areas above about $\pi$, the model is
in perfect agreement with the full calculation and with the experimental
results. For very high pulse areas calculations predict that the efficiency
should increase again because, like in the case of the reappearance of Rabi
rotations, the splitting between the dressed states becomes larger than the
cut-off frequency $\omega_c$ in the phonon spectral density
\cite{reiter2012pho}. With increasing temperatures the rate of phonon
absorption approaches the phonon emission rate and thus the asymmetry between
positive and negative chirp disappears \cite{luker2012inf,debnath2012chi}.
The ARP scheme then becomes less efficient except for a pulse area in a small
range around $\pi$ where phonon-induced transitions are not yet very
efficient (see Fig.~\ref{fig:arp_theo}(b)).

\begin{figure}[!ht]
\centering
\includegraphics[width=7cm]{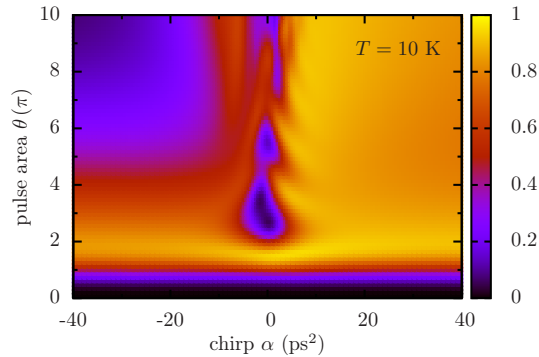}
\caption{\label{fig:arp_biex} Calculated biexciton occupation as a function of
pulse area and chirp coefficient for the case of
a two-color excitation using circularly polarized
pulses. The frequencies at the pulse maximum of the two pulses are taken to coincide
with the ground state-to-exciton and the exciton-to-biexciton transition, respectively.
Figure reprinted with permission from Gl{\"a}ssl \emph{et al} \cite{glassl2013bie}.}
\end{figure}

When exciting the QD with linearly polarized chirped pulses transitions to
the biexciton state become possible. In this case, also for positive chirps
the efficiency of exciton state preparation may be reduced due to a
phonon-assisted generation of the biexciton, which becomes important for
small biexciton binding energies \cite{gawarecki2012dep}. On the other hand,
this also opens up the possibility of a direct preparation of the biexciton
state using ARP, a process which is very attractive because of the high
potential use of biexciton states in quantum information technology
\cite{hui2008pro,glassl2013bie,debnath2013hig}. Two protocols either via
direct two-photon absorption resonant on the ground state-to-biexciton
transition or via two-color excitations have been proposed. In the first
case, a linearly polarized pulse is used to directly excite the biexciton
from the ground state. Because of the off-resonant coupling to the exciton
state higher pulse areas are required to reach the adiabatic regime. As soon
as this threshold is overcome, a stable biexciton generation via ARP is
possible, which can again be understood in the dressed state picture
\cite{glassl2013bie}. In the low temperature limit, the influence of phonons
is asymmetric with respect to the sign of the chirp in the same way as for
the exciton preparation. In the high temperature case, an overall damping is
found, but also for the biexciton preparation it is possible to decouple the
phonons from the electronic system by using pulses with a sufficiently high
intensity such that the splitting between the dressed states becomes larger
than the cut-off frequency for the phonon coupling. In this regime a robust
and high-fidelity biexciton preparation using ARP has been predicted at
temperatures as high as $80$~K \cite{debnath2013hig}. Instead of using a
linearly polarized pulse the two-color excitation scheme is based on the
excitation by a pair of chirped pulses with opposite circular polarization.
The frequency at the pulse maximum of one of the pulses is tuned to the
ground state-to-exciton and of the other to the exciton-to-biexciton
transition. Figure~\ref{fig:arp_biex} shows that this scheme again gives rise
to an efficient biexciton generation. The drawback of using two different
pulses is compensated by the fact that the adiabatic threshold lies at lower
pulse areas compared to the two-photon scheme \cite{glassl2013bie}.




\subsection{Phonon assisted exciton and biexciton preparation}
\label{sec:detuned}

\begin{figure}[!ht]
\centering
\includegraphics[width=5cm]{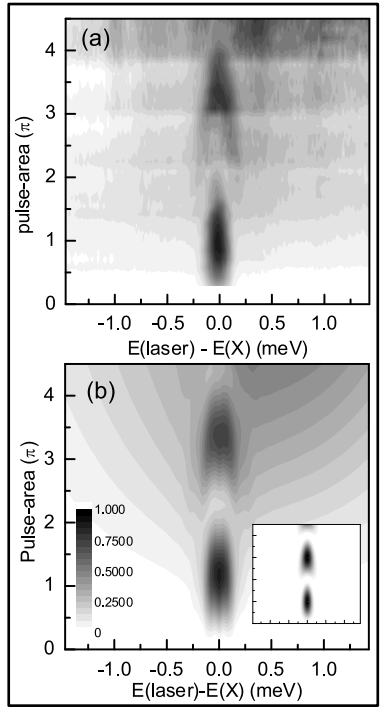}
\caption{\label{fig:ramsay}
(a) Gray scale plot of the photocurrent of a single QD excited
by a Gaussian pulse of  $4$~ps duration (full width at half maximum) at a
temperature of $15$~K measured as a function of laser detuning
and pulse area.
(b) Calculation of the photocurrent using a Markovian
master equation approach. (Inset) Calculation without dephasing.
Figure reprinted with permission from Ramsay \emph{et al} \cite{ramsay2011eff}.
}

\end{figure}

A Rabi-type dynamics is not restricted to the case of resonant driving but
can also be observed under off-resonant driving conditions. In an ideal TLS
without dissipation, off-resonant Rabi dynamics is characterized by
oscillations of small amplitude and high frequency [cf. Eq.~(\ref{eq:rabi})]
and not sensitive to the sign of the detuning \cite{allen1975opt}. This
behavior is significantly changed by the interaction with phonons. As in the
resonant case, the oscillations are damped and, for larger detunings, where
in the phonon-free case the amplitude is small, they are hardly resolved
\cite{ramsay2011eff,glassl2011lon,reiter2012pho,glassl2013pro}. In addition,
at low temperatures there is a strong asymmetry between positive and negative
detunings. For positive detunings, i.e., a laser frequency above the
resonance, superimposed on the weak oscillations there is a growth of the
exciton occupation as a function of time, whereas it decreases for negative
detunings \cite{glassl2011lon,hughes2013pho}. This asymmetry reflects a
thermalization in the dressed state basis and can be traced back to the fact
that immediately after switching on a pulse with positive detuning, the
dressed state with the higher ground state contribution is more strongly
occupied. For positive detunings this is the upper dressed state [cf.
Eq.~(\ref{eq:dressed-states})] while for negative detunings it is the lower
one. Thermalization at low temperatures is mainly governed by phonon emission
processes leading at longer times to a dominant occupation of the lower
dressed state, which for positive detunings has a large exciton state
contribution. Thus, for positive detunings the exciton state occupation
increases, while for negative detunings it remains small or even decreases.
With increasing temperature phonon absorption processes become more important
leading, like in the case of chirped pulse excitations, finally to an
equilibration between the dressed states and thus also between ground and
exciton state.

This asymmetry has also been demonstrated in experiments using pulsed
excitations \cite{ramsay2011eff}. Figure~\ref{fig:ramsay}(a) shows a
photocurrent spectrum of the QD occupation as a function of the detuning and
the pulse area. Clearly seen are Rabi oscillations for near resonant
excitations. The oscillation amplitude rapidly decays with rising detuning.
The main signature of the phonon-induced asymmetry is an enhanced exciton
occupation observed for positive detunings and higher pulse areas. Part (b)
of Fig.~\ref{fig:ramsay} displays results of corresponding calculations based
on a simple master equation approach that relies on Markovian rates for
transitions between the dressed states. Obviously, all key features of the
measurements are well reproduced on this level of the theory. Similar results
have also been obtained by calculations within a correlation expansion for
finite pulses \cite{reiter2012pho}. The calculations also reveal that the
asymmetry originates from the phonon influences, as can be seen, e.g., by a
comparison with simulations for the case without phonons shown in the inset
of Fig.~\ref{fig:ramsay}(b).

\begin{figure}[!ht]
\centering
\includegraphics[width=7cm]{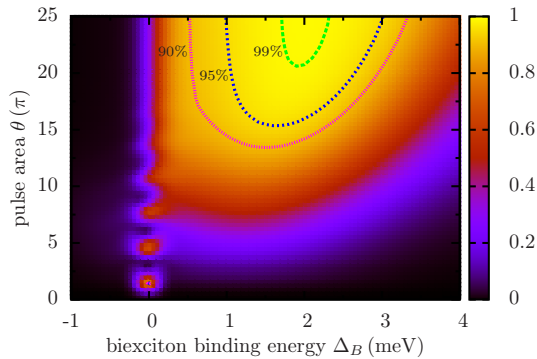}
\caption{\label{fig:bi-prep-offres}
Final biexciton occupation after a Gaussian pulse with a
FWHM of 15 ps at $T=4$~K as a function of the biexciton binding energy
$\Delta_B$ and the pulse area $\theta$.
The laser frequency is chosen in resonance with the ground state to exciton transition.
The contour lines display where certain values of the biexciton occupation are reached.
Figure reprinted with permission from Gl{\"a}ssl \emph{et al} \cite{glassl2013pro}.}
\end{figure}

The interpretation in terms of a relaxation between the dressed states
indicates that at low temperatures at sufficiently long times an essentially
complete occupation of the lower dressed state should be achievable. Since
for increasing positive detuning the lower dressed state becomes more and
more exciton-like, there should be no fundamental limitation in the final
exciton state occupation. Indeed, calculations for excitations with a
constant light field have shown that exciton occupations arbitrarily close to
one can be reached \cite{glassl2011lon}. However, in order to make
off-resonant excitations useful for state preparation purposes, the process
has also to be sufficiently fast such that other relaxation mechanisms not
included in the model, e.g., the radiative decay, are not yet efficient in
reducing the exciton occupation again. For dots without a cavity, this is not
a very severe restriction, as the typical time scale for radiative decay is
of the order of one nanosecond
\cite{borri2001ult,bayer2002tem,langbein2004rad}. For dots in cavities, often
cavity losses are the main relaxation process
\cite{reithmaier2004str,reithmaier2008str} which, depending on the quality
factor, can demand for noticeably shorter preparation times.

A short time scale of phonon-induced relaxation and a high fidelity of
exciton generation seem to be two conflicting ingredients. The phonon
emission rate is determined on the one hand by the phonon spectral density
[Eq.~(\ref{eq:phonon-spect})] at the energy of the splitting between the
dressed states, and on the other hand, since phonons only couple to the
excitonic part of the electronic state, by the exciton contribution in the
dressed states. This favors energy splittings in a certain energy window
(cf.~Fig.~\ref{fig:spectraldensity}) and parameters such that both dressed
states as given in Eq.~(\ref{eq:dressed-states}) have a noticeable exciton
contribution. To achieve a high fidelity, the lower dressed state that is
eventually reached at the end of the preparation process, should be as close
as possible to the exciton state. In the case of excitation by a rectangular
pulse, when the character of the dressed states remains constant over the
pulse, there is a conflict between these two requirements. The condition of a
strongly exciton-dominated lower dressed state means that the upper dressed
state almost coincides with the ground state giving rise to a very small
overlap and thus to a very low phonon emission rate. The minimal time
required to achieve a certain degree of fidelity is therefore determined by a
trade-off between these two opposite aspects. In the case of excitation by a
Gaussian pulse the character of the dressed states changes over the duration
of the pulse. Close to the pulse maximum a strong driving introduces a strong
mixing between ground and exciton state in both dressed states, while for
decreasing amplitude the lower dressed state evolves into the pure exciton
state thus favoring a high-fidelity preparation. Analogous arguments given
here for the exciton system hold for the exciton-biexciton system.

Recent calculations have shown that under realistic excitation conditions a
high-fidelity and robust phonon-assisted preparation targeted at either the
exciton or the biexciton is indeed possible on a picosecond time scale
\cite{glassl2013pro}. As an example, Fig.~\ref{fig:bi-prep-offres} displays
results for the preparation of the biexciton where the final biexciton
occupation after the pulse is plotted as a function of the biexciton binding
energy $\Delta_{B}$ and the pulse area $\theta$ at $T=4$~K. In these
calculations, a linearly polarized laser pulse with duration $15$~ps FWHM has
been tuned in resonance with the ground state-to-exciton transition. For
vanishing $\Delta_{B}$, the excitation is also resonant to the
exciton-to-biexciton transition, and two-photon Rabi oscillations between the
ground and the biexciton state are observed. For finite $\Delta_{B}$, in
turn, the exciton-to-biexciton transition is off-resonant and for positive
$\Delta_{B}$, i.e., when the biexciton energy is lower than twice the exciton
energy, the biexciton occupation increases reflecting the fact that the
lowest dressed state, towards which the system relaxes due to the
carrier-phonon interaction, has a dominant biexciton contribution. It turns
out that for a broad range of typical biexciton binding energies and
sufficiently high pulse areas an almost perfect preparation with biexciton
occupations close to one can be achieved. The phonon-coupling between the
dressed states turns out to be sufficiently strong, such that the final
values plotted in Fig.~\ref{fig:bi-prep-offres}, which document a preparation
of an almost pure biexciton state, are reached on a time scale of $\sim 10$
ps \cite{glassl2013pro}. In order to realize such short preparation times,
one exploits the fact that the phonon-induced relaxation strongly depends on
the system parameters such as the QD size and the electron-phonon coupling
strength as well as on the driving parameters such as the pulse area and the
pulse length. Once the parameters are chosen in a regime favorable for fast
relaxation, the protocol turns out to be robust with respect to changes in
the pulse area as well as to the precise value of the biexciton binding
energy.

Being caused by phonon emission processes, the state preparation speeds up,
when the carrier-phonon coupling increases. Like all the other results
discussed in this review, the calculations shown in
Fig.~\ref{fig:bi-prep-offres} correspond to the weak coupling limit realized
for InGaAs-type QDs, while for other materials such as, e.g., GaN, the
carrier-phonon coupling can be substantially stronger
\cite{krummheuer2005pur,ostapenko2012exc}. Thus, in contrast to the other
preparation protocols discussed in this review, which work best in the
absence of the carrier-phonon coupling, the phonon-assisted preparation based
on off-resonant driving performs the better the stronger the carrier-phonon
coupling is \cite{glassl2013pro}. Finally, we note that the high biexciton
occupations found in Fig.~\ref{fig:bi-prep-offres} also imply that the
exciton state is eventually unoccupied even though it is resonantly driven.
This is because the lowest dressed state has an almost vanishing exciton
contribution. In fact, the resonant driving of the exciton with linearly
polarized laser pulses turns out to be most favorable for preparing the
biexciton. Using lasers that are off-resonant to all transitions in the
exciton-biexciton system is typically less efficient.

\begin{figure}[!ht]
\centering
\includegraphics[width=7cm]{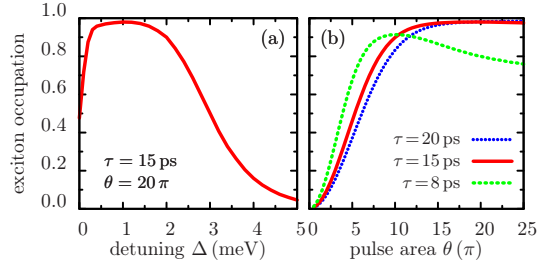}
\caption{\label{fig:x-prep-offres}
Final exciton occupation after a Gaussian pulse with pulse
length $\tau$ (FWHM) at $T=4\,\rm{K}$, (a) as a function
of $\Delta$ for $\tau=15\,\rm{ps}$ and $\theta=20\pi$, and (b) as a
function of $\theta$ for $\Delta=1\,\rm{meV}$ and different values of
$\tau$ as indicated.
Figure reprinted with permission from Gl{\"a}ssl \emph{et al} \cite{glassl2013pro}.}
\end{figure}

Similar to the biexciton state preparation also the single exciton state can
be efficiently populated by a phonon-assisted process \cite{glassl2013pro}.
This is achieved by a circularly polarized excitation -- in order to suppress
the coupling to the biexciton -- by a laser pulse which is positively detuned
with respect to the ground state-to-exciton transition. For such an
excitation, Fig.~\ref{fig:x-prep-offres}(a) illustrates the dependence of the
exciton occupation after the pulse on the detuning $\Delta$. For small and
large $\Delta$, low exciton occupations are found, while high occupations are
obtained for intermediate $\Delta$, reflecting the finite energy window of
efficient carrier-phonon coupling in the phonon spectral density
(cf.~Fig.~\ref{fig:spectraldensity}). In Fig.~\ref{fig:x-prep-offres}(b) the
final exciton occupation is plotted as a function of the pulse area for
several pulse lengths. Clearly, when the pulse length is below 10 ps, the
maximal occupation that can be reached is noticeably lower than one and
quickly further degrades when the pulse area is not optimally chosen.
However, already for pulse durations above $15$~ps, the final exciton
occupation stays close to one for a broad range of pulse areas.
Figure~\ref{fig:x-prep-offres} therefore again confirms the robustness of the
phonon-assisted scheme against variations of detuning, pulse area and pulse
length within certain windows determined by material and excitation
parameters.

\section{Conclusions}
\label{sec:conclusions}

In this review we have discussed various schemes that have been used for the
preparation of excitonic and biexcitonic states in self-assembled QDs by
means of optical excitations. The focus of the discussion was on the role of
phonons for these excitation schemes. Three basically different schemes have
been presented in the previous section: (i) an excitation by resonant laser
pulses employing Rabi oscillations, (ii) an excitation by using chirped laser
pulses based on ARP, and (iii) an excitation by detuned laser pulses relying
on phonon-assisted exciton or biexciton generation. In the following we will
compare these schemes regarding different aspects and discuss the advantages,
drawbacks, and limitations of the respective scheme.

First of all, when talking about state generation one has to distinguish
between the generation of pure exciton or biexciton states and the generation
of arbitrary superposition states. Of the three schemes discussed in this
review only the resonant excitation can be used for the preparation of
superposition states. Superposition states are in general subject to
phonon-induced dephasing after their generation. However, due to the
non-Markovian nature of the pure dephasing process this dephasing is not
complete. By excitation with not too short laser pulses the irreversibility
which is associated with the emission of a phonon wave packet can be reduced.
For long pulses, on the other hand, the dephasing during the excitation
process caused by phonon-induced transitions between the dressed states as
well as by other dephasing mechanisms becomes more important, such that an
optimum for the pulse duration exists. The fidelity of the state preparation
in this case strongly depends on the details of the QD structure and on the
temperature.

Pure exciton and biexciton states can be prepared by all three mechanisms.
For all three schemes there is no fundamental limitation of the achievable
fidelity caused by the interaction with phonons, although limitations may
arise from the approximations inherent in the model, e.g., by neglecting
higher excitonic states as well as other coupling mechanisms, in particular
the radiative decay. In the case of Rabi rotations the fidelity of exciton or
biexciton preparation can be made arbitrarily high in the limit of ultrafast
excitations, because here the dynamics in the exciton system becomes much
faster than the phonon dynamics such that the phonons cannot follow. In the
ARP scheme using excitations with chirped pulses a high-fidelity generation
of excitons and biexcitons can be reached. In the case of positive chirps at
sufficiently low temperatures, when phonon absorption processes are
negligible, a high-fidelity preparation is possible in a wide range of pulse
areas while for negative chirps the pulse areas are restricted to a window
between about $\pi$ and $2\pi$. For excitations with detuned pulses exciton
and biexciton occupations arbitrarily close to one can be achieved by using
sufficiently long pulses at low temperatures such that the relaxation from
the upper to the lower dressed state by phonon emission is essentially
complete and phonon absorption processes can be neglected.

A clear advantage of schemes (ii) and (iii) over the resonant excitation
scheme (i) is their robustness with respect to pulse and system parameters.
Once a certain threshold of pulse intensity and pulse duration is overcome,
the exciton or biexciton generation is rather insensitive to variations of
these parameters. This has the advantage that no detailed knowledge of the
individual QD is required to perform the state preparation. Furthermore, for
this reason the schemes also work for ensembles of QDs with a distribution of
energies and dipole matrix elements, as long as these distributions are
sufficiently narrow. In the case of ARP the adiabaticity condition has to be
fulfilled for the whole ensemble, in the case of off-resonant excitation the
splitting of the dressed states of each QD in the ensemble has to be in the
range such that phonon emission occurs with a sufficient rate and phonon
absorption processes are negligible.

When comparing the robust schemes (ii) and (iii) one observes that for
typical QD structures ARP works often best for pulse areas given by a few
multiples of $\pi$ while the off-resonant preparation requires pulse areas
above $10\pi$. On the other hand, they have an opposite trend when varying
the phonon coupling constant, e.g., by considering different material
systems. Since phonon emission becomes more efficient for increasing
exciton-phonon coupling the pulse area required for the state preparation in
scheme (iii) decreases, while the coherent schemes (i) and (ii) typically
deteriorate for increasing phonon coupling.

Also the temperature plays a quite different role in the different schemes.
Schemes (ii) and (iii) rely on the absence of phonon absorption processes
between the dressed states. Since the splitting between the dressed states is
in the range of a few meV, already at temperatures of about $20$~K the rate
of phonon absorption is of the same order as the phonon emission rate.
Therefore, an equal occupation of ground and exciton state will be reached
and the state preparation becomes rather inefficient. In the case of ARP this
limitation may be partly overcome either by using pulses with rather low
pulse areas around $\pi$, where phonon-induced transitions are not yet very
efficient or by using pulses with very high pulse areas, where the phonons
are effectively decoupled. In the resonant scheme (i), on the other hand, at
least a few Rabi rotations have been clearly observed up to temperatures of
$50$~K. Even higher temperatures should in principle be possible by using
shorter pulses, as long as the increased spectral width of the pulse does not
lead to the excitation of other nearby transitions which may be present in a
real QD.

The above discussion clearly demonstrates that there is not a single simple
criterion which defines the ``best'' state preparation scheme in a QD.
Instead, the selection of the optimal scheme for a given purpose depends on
the final state to be prepared as well as on a variety of different
conditions such as material parameters or temperature, but also on
limitations due to phenomena not included in the model. Other nearby
transitions may set a lower limit for the pulse duration, while additional
dephasing or relaxation channels may set an upper limit for the time of the
preparation process. However, the discussion also clearly demonstrates that
the underlying model, despite its simplicity, is able to quantitatively
describe a variety of experimental results and therefore can be seen as a
prototypical model for an interacting many-body system which, due to its
simplicity, is able to provide detailed physical insights in the dynamics of
coupled fermion-boson systems.

\ack This work was supported by the Deutsche Forschungsgemeinschaft DFG
through the grant Ax 17/7-1.


\providecommand{\newblock}{}

\end{document}